\documentclass[12pt,longnamesfirst,preprint]{aastex}

\slugcomment{2010 Asia-Pacific Radio Science Conference\\
 in Toyama, Japan on September 22-26, 2010\\
Commission J: Radio Astronomy\\
J2 Millimeter- and Sub-millimeter-wave Telescope and Array 
}

\shorttitle{Saito et al.}
\shortauthors{
Atacama Compact Array Antennas
}

\begin{document}


\title{
Atacama Compact Array Antennas
}


\author{
Masao Saito\altaffilmark{1},
Inatani, Junji\altaffilmark{1}, Nakanishi Kouichiro\altaffilmark{1}, Naoi Takahiro\altaffilmark{1}, Yamada Masumi\altaffilmark{1}, Hiro Saito\altaffilmark{1}, Bungo Ikenoue\altaffilmark{1}, Yoshihiro Kato\altaffilmark{1}, Kou-ichiro Morita\altaffilmark{2}, Norikazu Mizuno\altaffilmark{1}, and Satoru Iguchi\altaffilmark{1}}  


\altaffiltext{1}{
National Astronomical Observatory of Japan, 2-21-1 Osawa,
Mitaka, Tokyo 181-8588, Japan
}
\altaffiltext{2}{
Joint ALMA Office, Santiago Chile
}

%\begin{abstract} 

%\end{abstract}

%\keywords{} 

\section{ALMA and ACA} 
ALMA (Atacama Large Millimeter/submillimeter Array) is a gigantic radio interferometer array with 66 parabola antennas under construction in the Atacama Desert at an altitude of approximately 5000 meters in northern Chile [1] and its regular operation will start from 2012 in the frequency range from 31.3 to 950 GHz (10 $-$ 0.35 millimeter in wavelength). ALMA consists of fifty 12-m antennas and "Atacama Compact Array (ACA)". The ACA is composed of four 12-m antennas and twelve 7-m antennas both of which are delivered by NAOJ. By spreading these transportable antennas over the distance of up to 18.5 kilometer, ALMA achieves the angular resolution equivalent to a telescope of 18.5 kilometer in diameter, as a telescope with the world's highest sensitivities and angular resolutions at millimeter and submillimeter wavelengths. The ACA system aims to acquire the total-power and short-baseline interferometer data that cannot be taken with the array of only 12-m antennas, which increases reliability of the interferometer maps of astronomical sources larger than the field view of the 12-m antenna. 

\section{ACA Antenna Design and Requirements}
Both 12-m and 7-m antennas will are constituted by a symmetrical paraboloidal reflector with a Cassegrain optical layout mounted on an Altitude-Azimuth mount. The primary reflector surface consists of machined aluminum panels with a suitable surface finish to enable direct solar observing. The reflector surface is mounted by means of adjusters onto a reflector backup structure (BUS). The subreflector with its mechanism is supported by feed legs in a quadripod configuration. The position of the subreflector is remotely adjusted with a controlled mechanism for focusing and collimation. The online pointing correction system was introduced to satisfy a stringent requirement of pointing and tracking performances. 
In order to achieve high sensitivity and high image quality, the requirement of the ACA antennas are very challenging. The four major performances are all-sky pointing (to be not more than 2.0 arcsec), offset pointing (to be $<$ 0.6 arcsec) surface accuracy ($<$ 25 micrometer), stability of path-length (15 micrometer over 3 min), and high servo capability (6 degrees/s for Azimuth and 3 degrees/s for Elevation) under the realistic daytime and nighttime conditions of the Array Operation Site (AOS) at an altitude of approximately 5000 meters.

\section{Measured Major Performance}  
The high performance of the ACA antenna has been extensively evaluated at the Site Erection Facility (SEF) area in the Operations Support Facility (OSF) at an altitude of about 2900 meters. Pointing performance was evaluated with a specially dedicated Optical-Pointing Telescope (OPT) aboard the ACA antenna. Near field holography was used to set and measure the reflector surface under various weather conditions. Fast motion capabilities were tested using encoder readings. Test methods (e.g. [2]) and results will be presented in the poster.

\end{document}